\documentclass[11pt]{article}
\usepackage{amsmath,amssymb,amsthm,easybmat,verbatim,float}
\usepackage{enumerate}
\usepackage{color}
\usepackage{bm}
\usepackage{graphicx}
\usepackage{longtable}
\usepackage{array}
\newcolumntype{x}[1]{%
>{\centering\arraybackslash}p{#1}}%

\newcommand{\sepAuthor}{0.5in}
\newcommand{\sepAbstract}{0.4in}
\newcommand{\skipKeywords}{30pt}

\long\def\mytitlepage#1#2#3#4{
        \thispagestyle{empty}
        \begin{center}
        {\Large\bf #1}

        \vspace{\sepAuthor}
        #2\\
        \medskip

        \vspace{\sepAbstract}
        {\Large Abstract}
        \end{center}

        \noindent{#3}
        \vskip\skipKeywords

        \noindent{#4}
        \clearpage
        }

\newtheorem{lemma}{Lemma}

\newtheorem{theorem}{Theorem}

\newcommand{\bra}[1]{\langle #1|}
\newcommand{\ket}[1]{|#1\rangle}

\newcommand{\op}[2]{|#1\rangle \langle #2|}
\newcommand{\idmap}{{\mathbb{I}}}

\newcommand{\SLOCCprec}{\le_{\textrm{\tiny SLOCC}\!}}
\newcommand{\LOCCprec}{\le_{\textrm{\tiny LOCC}\!}}

\begin{document}
\mytitlepage{Matrix Pencils and Entanglement Classification\!
\footnote{This work was supported in part by the National Science
Foundation of the United States under Awards~0347078 and 0622033.
}}{
Eric Chitambar\\
Physics Department, University of Michigan\\
450 Church Street, Ann Arbor, Michigan 48109-1040, USA\\
E-mail: echitamb@umich.edu\\
\vspace{2ex}
Carl A. Miller\\
Department of Mathematics, University of Michigan\\
530 Church Street,  
Ann Arbor, MI 48109-1043, USA\\
E-mail: carlmi@umich.edu\\
\vspace{2ex}
Yaoyun Shi\\
Department of Electrical Engineering and Computer Science\\
University of Michigan\\ 
2260 Hayward Street, Ann Arbor, MI 48109-2121, USA\\
E-mail: shiyy@umich.edu
}{\noindent 
Quantum entanglement plays a central role in quantum information processing.
A main objective of the theory of quantum entanglement is to classify different types of entanglement
according to their inter-convertibility through manipulations that do not require quantum communication.
While bipartite entanglement is well understood in this framework, 
the classification of entanglements among three or more subsystems is inherently much more difficult. 

In this paper, we study pure state entanglement in systems of dimension $2\otimes m\otimes n$.  Two states are considered equivalent if they can be reversibly converted from one to the other with a nonzero probability using only local quantum resources and classical communication (SLOCC).  We introduce a connection between entanglement manipulations in these systems and the well-studied theory of matrix pencils.  All previous attempts to study general SLOCC equivalence in such systems have relied on somewhat contrived techniques which fail to reveal the elegant structure of the problem that can be seen from the matrix pencil approach.  Based on this method, we report the first polynomial-time algorithm for deciding when two $2\otimes m\otimes n$ states are SLOCC equivalent.  Besides recovering the previously known 26 distinct SLOCC equivalence classes in $2\otimes 3\otimes n$ systems,
we also determine the hierarchy between these classes.
}{}

\section{Introduction}
The feature that most distinguishes multipartite quantum systems from their classical counterpart is their ability to be in so-called entangled states.  Not only does quantum entanglement enable apparent ``spooky action at a distance'' between separated systems \cite{Einstein-1935a}, but it also has the potential to fundamentally change and dramatically improve the current information processing and cryptographic technologies \cite{Mermin-2007a}.  It becomes natural then to treat entanglement in a multipartite system as a information
processing resource that takes on different forms as the system realizes different states.  Much effort has been devoted to formally quantifying the amount of entanglement contained in a given quantum state with the motivating intuition being that states having more entanglement possess a greater degree of computational and communication power than those with a lesser amount.  
   
Under this interpretation, one may reasonably argue that a state $\ket{\phi}$ possesses no less amount of entanglement than another
state $\ket{\psi}$ of the same system if the system can be converted from $\ket{\phi}$ to $\ket{\psi}$
``free of charge,'' or without needing any further entanglement to facilitate the transformation.
The physical operations implementing such transformations is the celebrated class of Local Operations with Classical Communication (LOCC), 
which, as described by its name, consists of all operations in which each multipartite subsystem is manipulated locally but perhaps in a manner globally correlated through classical communication.  Thus LOCC has become a major framework for quantifying and classifying entanglement (see, e.g., the recent
surveys on quantum entanglement by Horodecki {\em et al.}~\cite{Horodecki} and G{\"u}hne and T{\'o}th~\cite{Guhne}).

Following the notation of Bennett {\em et al.}~\cite{Bennett_SLOCC}, we write $\ket{\psi}\LOCCprec\ket{\phi}$ if $\ket{\phi}$ can be converted to $\ket{\psi}$
through LOCC. When can a state be converted to another? What's the largest ratio one can convert multiple copies of a state to multiple copies of another? When is there a state maximum in the sense
that all other states in the systems can be obtained from this state? Those are examples of the many natural
questions that arise.  We often have answers for the {\em bipartite} case.
For example, a remarkable theorem by Nielsen~\cite{Nielsen-1999a} states that for {\em bipartite} states, $\ket{\psi}\LOCCprec\ket{\phi}$ if and only
if the spectrum of $\ket{\phi}$'s reduced density operator (i.e. the {\em Schmidt numbers})
is majorized by that of $\ket{\psi}$.  Another important observation, made by
Bennett {\em et al.}~\cite{bennett-1999c}, is that (for any multipartite systems) if two states are equivalent under LOCC, they are related by a local unitary (LU) transformation. Thus LOCC equivalence classes are simply the orbits of local unitary operations. Such a partitioning is too fine for most interests: even in the two-qubit case, there exists an infinite number of LU equivalence classes.  

However, if the required success probability of both the forward and reverse transformations is reduced to be simply nonzero, a much coarser partitioning is achieved.  General LOCC transformations occurring with a nonzero probability are called stochastic (SLOCC) and denoted by $\ket{\psi}\SLOCCprec\ket{\phi}$ if the transformation
is from $\ket{\phi}$ to $\ket{\psi}$. It turns out that SLOCC equivalence classes are precisely the orbits under local invertible linear transformations~\cite{Dur-2000a}.
 Similar to the situation with LOCC, bipartite entanglement is well understood under SLOCC. Indeed, for bipartite pure states, $\ket{\psi}\SLOCCprec\ket{\phi}$
if and only if the rank of the reduced density operator of $\ket{\psi}$ (i.e. the Schmidt rank) is no larger than that of $\ket{\phi}$. Thus two states are SLOCC equivalent if and only if they 
have the same Schmidt rank. The optimal success probability can also be computed easily from the Schmidt numbers~\cite{Vidal-1999a}.  

In contrast, entanglement among three or more parties behaves fundamentally different from bipartite entanglement. For example, while there is a maximum SLOCC equivalence class
for bipartite systems of any dimension, there exists two maximal equivalence classes for the simplest tripartite system
of $3$ qubits~\cite{Dur-2000a}. 
In contrast to Nielsen's Theorem and the rank criterion for bipartite SLOCC conversion, deciding SLOCC convertibility in general encodes many difficult computational problems.
For the general tripartite conversions, the problem is NP-hard (observed in~\cite{Chitambar-2008a} using a NP-hardness
result by H{\aa}stad~\cite{Haastad-1990a} on computing tensor rank). For converting a tripartite state to a bipartite state, it is equivalent to 
the important problem of Polynomial Identity Testing~\cite{Chitambar-PIT}. 
For certain tripartite asymptotic conversion the optimal
conversion ratio is precisely the exponent of matrix multiplication~\cite{Chitambar-2008a}. In view of those results, 
a simple criterion or an efficient algorithm for checking SLOCC convertibility or equivalence 
could be found only for systems of restricted dimensions.

This article studies the SLOCC equivalence classes of tripartite pure states in systems of dimensions $2\otimes m\otimes n$. 
Dur \textit{et al.} presented the first major result in the study of multipartite SLOCC equivalence classes by showing there to be six different classes in $2\otimes 2\otimes 2$ systems \cite{Dur-2000a}.  Their work was extended to four qubit systems by Verstraete and co-workers in which already an infinite number of equivalence classes exist \cite{Verstraete-2002a}.  For an arbitrary number of subsystems, Miyake has shown how multidimensional determinate theory can be used to obtain general properties and results concerning SLOCC equivalence \cite{Miyake-2003a}.  Specific to tripartite $2\otimes 2\otimes n$ systems, Miyake and Verstraete have also completely characterized the equivalence class hierarchy and found that for $n\geq 4$ exactly nine different classes exist \cite{Miyake-2004a}.  Using the method of successive Schmidt decompositions, Cornelio and Piza obtained partial results concerning the equivalence classes in $2\otimes m\otimes n$ systems \cite{Cornelio-2006a}. Chen \textit{et. al}  completed the finite orbit picture by enumerating all 26 equivalence classes in $2\otimes 3\otimes 6$ systems, and showed that for the $3\otimes3\otimes3$ and $2\otimes4\otimes4$ systems (and all systems of higher dimensions) there are infinite number of SLOCC equivalence classes.
They used a technique called ``the range criterion'' \cite{Chen-2006a}, which states that two states are SLOCC equivalent if and only if
the ranks of the reduced density operators are identical and their supports are related by local invertible linear operations. 
While these results are quite interesting, the tools used to obtain them appear rather \textit{ad hoc} and neither the criterion in Ref \cite{Chen-2006a} nor any previous technique provides an efficient algorithm (or any algorithm at all) for determining SLOCC equivalence.  The non-invertible hierarchy among the 26 classes has also remained an open problem.

\begin{comment}
As for transformations between inequivalent classes, very little is known in the general multipartite setting.  The bipartite picture is complete with necessary and sufficient conditions determined for both probabilistic and deterministic transformations \cite{Vidal-1999a, Nielsen-1999a}.  Ref. \cite{Miyake-2004a} provides the largest known generalization to multipartite systems in which the authors provide a complete hierarchy between the nine equivalence classes thus enabling a method for determining whether $\ket{\phi}\SLOCCprec\ket{\psi}$ in $2\otimes 2\otimes 4$ systems. 
\end{comment}

The main insight of this article is that the theory of matrix pencils is the perfect tool for analyzing SLOCC equivalence in $2\otimes m\otimes n$ systems.
For two matrices $A, B\in\mathbb{C}^{m\times n}$, the linear matrix polynomial $\lambda A + \mu B$ is called a matrix pencil. Two pencils $\lambda A+\mu B$ and
$\lambda A'+ \mu B'$ are equivalent if there exists invertible $P$ and $Q$ such that $P(\lambda A+\mu B)Q = \lambda A' + \mu B'$. A fundamental result is the existence of a canonical form, discovered by Kronecker (see, e.g., Gantmacher~\cite{Gantmacher-1959a}). The theory of matrix pencils remained an important 
subject of study for its applications in control and systems theory. An example is the computation of the generalized eigenvalues (see, e.g., Section 7.7 of \cite{Golub}).\footnote{For two matrices $A$ and $B$, a vector $x$ and a constant $\lambda$, if $Ax=\lambda Bx$, $\lambda$ is a generalized eigenvalue of $(A, B)$ and $\lambda$ the associated eigenvector. The set of generalized eigenvalues
are precisely $\{\lambda/\mu: \det(\lambda A+ \mu B)=0\}$.}  The efficient computation of the Kronecker canonical forms, other canonical forms, and related problems is still
an active field of research (see, e.g. \cite{Beelen} and following articles).

The connection with our problem is that each state in a $2\otimes m\otimes n$
space can be represented as a matrix pencil (see Section~\ref{sec:connection} for details). The local operations on the second and third
subsystem brings the corresponding pencil to an equivalent one. While actions on the first subsystem may bring the pencil to an inequivalent one, we show that
if two states are SLOCC equivalent, there are only a small number of choices for operations on the first subsystem that make the resulting pencils equivalent.
As a consequence, we derive the first efficient algorithm (in fact, the first algorithm) for determining SLOCC equivalence in general $2\otimes m\otimes n$ systems. 
For the systems having a finite equivalence classes, we derive the equivalence orbits (which was known before) and represent them using Kronecker canonical forms.
We also determine all possible non-invertible transformations among those equivalence classes. 

The rest of this article begins with a brief introduction to some main results in matrix pencil theory.  We then develop the relationship between tripartite pure states and matrix pencils which allows us to derive necessary and sufficient conditions for the SLOCC convertibility of $2\otimes m\otimes n$ states.  From there, we develop a SLOCC hierarchical schematic of all tripartite systems possessing a finite number of SLOCC orbits.  The article closes with some brief concluding remarks.

%\bigskip\noindent
\section{Matrix Pencils}

The theory of matrix pencils was first developed by Kronecker over a century ago.  A completely thorough treatment of the subject can be found in Gantmacher's two volume texts \cite{Gantmacher-1959a} from which we will here only cite the main definitions and results.  For a more modern treatment, see Ref. \cite{Gohberg-2009a}.  Given two complex $m\times n$ matrices $R$ and $S$, we form the homogeneous matrix polynomial $\mathcal{P}_{(R,S)}=\mu R +\lambda S$ in variables $\mu$ and $\lambda$.  Two pencils $\mathcal{P}_{(R,S)}$ and $\mathcal{P}_{(R',S')}$ are \textbf{strictly equivalent} if there exists invertible matrices $B$ and $C$ independent of $\mu$ and $\lambda$ such that $\mu R'+\lambda S'=B(\mu R+\lambda S)C^T$.  It immediately follows that $\mathcal{P}_{(R,S)}$ and $\mathcal{P}_{(R',S')}$ are strictly equivalent if and only if there exists invertible $B$ and $C$ such that $BRC^T=R'$ and $BSC^T=S'$.  

The \textbf{rank} of $\mathcal{P}_{(R,S)}$ is the largest $r$ such that there exists an $r$-minor of $\mathcal{P}_{(R,S)}$ not identically zero (not equaling zero upon any complex substitution for $\mu$ and $\lambda$).  For $i\leq r$  we let $D_i(\mu,\lambda)$ denote the greatest common divisor of every $i$-minor of $\mathcal{P}_{(R,S)}$ which is monic with respect to $\lambda$.  The \textbf{invariant polynomials} of pencil $\mathcal{P}_{(R,S)}$ are the homogeneous polynomials $E_i(\mu,\lambda)=\frac{D_i(\mu,\lambda)}{D_{i-1}(\mu,\lambda)}$ for $i=1...r$ where $D_0(\mu,\lambda)\equiv 1$.  There will be a unique factorization of $D_r(\mu,\lambda)$ as $D_r(\mu,\lambda)=\mu^{r-k}p_1\cdot...\cdot p_k$ where $p_j$ is of the form $\mu x_j+\lambda$ for $x_i\in\mathbb{C}$, and the invariant polynomials will likewise have a factorization in terms of the $p_i$ and powers of $\mu$.  Then for each distinct $p_i$ (suppose there are $q\leq k$ of them) we can consider the ordered set $\{p_i^{e_{i1}},...,p_i^{e_{ir}}\}$ where $e_{ij}$ gives the largest power of $p_i$ that divides invariant polynomial $E_j$.  The multiset generated by letting $i$ range from 1 to $q$ is called the \textbf{elementary divisors} of $\mathcal{P}_{(R,S)}$.  Likewise, the set $\{\mu^{e'_1},...,\mu^{e'_r}\}$ is called the \textbf{infinite elementary divisors} of $\mathcal{P}_{(R,S)}$ where $e'_i$ is the largest power of $\mu$ that divides $E_i(\mu,\lambda)$.  From knowing the infinite and elementary divisors of $\mathcal{P}_{(R,S)}$, its invariant polynomials can be constructed and vice versa.

The right null space of $\mathcal{P}_{R,S}$ is the set of homogeneous polynomial vectors 
\[\textbf{x}_i(\mu,\lambda)=\sum_{j=0}^{\epsilon_i}x_{ij}\mu^{\epsilon_i-j}\lambda^j\] such that $(\mu R+\lambda S)\textbf{x}_i(\mu,\lambda)\equiv 0$ and $x_{i\epsilon_i}\not =0$.  A basis for the right null space whose elements have degrees $\epsilon_1\leq...\leq\epsilon_p$ is called \textbf{fundamental} if any other basis whose elements have degrees $\epsilon'_1\leq...\leq\epsilon'_p$ implies $\epsilon_i\leq\epsilon'_i$ for all $i$.  A important property of any $\textbf{x}_i(\mu,\lambda)=\sum_{j=0}^{\epsilon_i}x_{ij}\mu^{\epsilon_i-j}\lambda^j$ belonging to a fundamental set is that \textit{the $x_i$ are linearly independent} \cite{Gantmacher-1959a}.  Likewise, we can define the left null space to be homogeneous polynomial vectors satisfying $\mathcal{P}_{R,S}^T\textbf{x}_i(\mu,\lambda)=0$ and form fundamental sets having degrees $\nu_1\leq...\leq\nu_q$.  The values $\epsilon_1,...,\epsilon_p$ and $\nu_1,...,\nu_q$ are called the \textbf{minimal indices} of $\mathcal{P}_{R,S}$.  In particular, the number of $\epsilon_i$ that are zero will be called the \textbf{zero index number}, and the number of $\nu_i$ that are zero will be called the \textbf{transpose zero index number}.  With this overview, we can now state the main theorem characterizing strictly equivalent pencils.

\begin{lemma}[Kronecker]
\label{kron}
Two matrix pencils are strictly equivalent if and only if they have the same elementary divisors (finite and infinite) and the same minimal indices.  Moreover, suppose $\mathcal{P}_{(R,S)}$ has finite elementary divisors $\{(\mu x_1+\lambda)^{e_{11}},(\mu x_1+\lambda)^{e_{12}},...,(\mu x_q+\lambda)^{e_{qr}}\}$, infinite elementary divisors $\{\mu^{e'_1},...,\mu^{e'_r}\}$, minimal indices $\epsilon_1,...,\epsilon_p$ and $\nu_1,...,\nu_q$, a zero index number of $g$, and a transpose zero index number of $h$.  Then $\mathcal{P}_{(R,S)}$ is strictly equivalent to the canonical block-form diagonal pencil 
\begin{equation}
\label{eq:kron}
\{0^{h\times g},L_{\epsilon_{g+1}},...,L_{\epsilon_p},L^T_{\nu_{h+1}},...,L^T_{\nu_q},J\}
\end{equation}
where $0^{h\times g}$ is the $h\times g$ zero matrix, $L_\epsilon=\overbrace{\begin{pmatrix}\lambda&\mu&0&...&0\\0&\lambda&\mu&...&0\\&...\\0&0&...&\lambda&\mu \end{pmatrix}}^{\epsilon+1} \Bigg \}\epsilon\;$, and $J$ full rank square pencil with block-form 
\begin{equation}
\{N^{e'_1},...,N^{e'_r},M^{e_{11}},...,M^{e_{qr}}\}
\end{equation}
where $N^{e'_i}=\mu{\idmap}^{e'_i}+\lambda H^{e'_i}$ and $M^{e_{ij}}=(\mu x_i+\lambda){\idmap}^{e_{ij}}+\mu H^{e_{ij}}$ with ${\idmap}^t$ the $t\times t$ identity matrix and $H^t$ a $t\times t$ matrix whose only nonzero elements are ones on the superdiagonal.
\end{lemma}
\noindent We close this section by noting that both the minimal indices and elementary divisors of a pencil can be determined by efficient algorithms.  The first involves determining the null space of scalar matrices, and the latter amounts to performing Gaussian elimination on the matrix $\mu R+\lambda S$ \cite{Gantmacher-1959a}.

%\bigskip\noindent
\section{Connection to $2\otimes m\otimes n$ Pure States}
\label{sec:connection}

Any $2\otimes m\otimes n$ state can be expressed in bra-ket form as $\ket{\Psi}=\ket{0}_A\ket{R}_{BC}+\ket{1}_A\ket{S}_{BC}$.  By choosing local bases $\{\ket{i}_B\}_{i=0...m-1}$ and $\{\ket{i}_C\}_{i=0...n-1}$ for Bob and Charlie respectively, we can express the state as
\begin{equation}
\ket{\Psi}=\bigg(\ket{0}_A(R\otimes{\idmap})+\ket{1}_A(S\otimes{\idmap})\bigg)\ket{\Phi_n}=\bigg(\ket{0}_A({\idmap}\otimes R^T)+\ket{1}_A({\idmap}\otimes S^T)\bigg)\ket{\Phi_m}
\end{equation}
where $R_{ij}=\alpha_{ij}$, $S_{ij}=\beta_{ij}$, and $\ket{\Phi_k}=\sum_{i=0}^{k-1}\ket{i}_B\ket{i}_C$.  Thus, there is a one-to-one correspondence between a $2\otimes m\otimes n$ pure state $\ket{\Psi}$ and the pair of matrices $(R,S)$, so that to every $\ket{\Psi}$ and choice of indeterminates $\mu,\lambda$, we can uniquely associate the pencil $\mathcal{P}_{(R,S)}$ which we shall equivalently denote as $\mathcal{P}_{\Psi}$.  

There exists a nice relationship between the structure of $\mathcal{P}_{\Psi}$ and the local ranks of each subsystem.  The reduced states of Bob and Charlie are obtained by performing a partial trace on the matrix $\op{\Psi}{\Psi}$.  From above, then, it follows that
\begin{align}
\label{eq:reduced states}
\rho_B&=tr_{AC}(\op{\Psi}{\Psi})=RR^\dagger+SS^\dagger\notag\\
\rho_C&=tr_{AB}(\op{\Psi}{\Psi})=R^T\bar{R}+S^T\bar{S}.
\end{align}
Here, ``{\scriptsize T}'' denotes the matrix transpose with respect to the basis $\ket{i}_B{}_C\bra{j}$ and ``{\scriptsize -}'' the complex conjugate of its entries.  Also note that since Alice has a two dimensional system, her subsystem will either have full rank or be completely separated from Bob and Charlie.  Combining these facts, we can prove the following.
\begin{lemma}
\label{thm:localranks}
(i)  Bob and Charlie share pure entanglement (Alice separated) if and only if $\mathcal{P}_{\Psi}$ can be expressed as a matrix polynomial in one indeterminate $\hat{\lambda}$; i.e.
\begin{equation}
\mu R + \lambda S=\hat{\lambda}\hat{S},
\end{equation}
and (ii) Bob and Charlie's local ranks are $m-h$ and $n-g$ respectively where $g$ is the zero index number of $\mathcal{P}_{\Psi}$ and $h$ its transpose zero index number.
\end{lemma}
\begin{proof}[\bf Proof]
(i)  Alice is unentangled if and only if up to an overall phase, the state can be written as $\ket{0}(\ket{R}+\ket{S})+\alpha\ket{1}(\ket{R}+\ket{S})$ which happens if and only if its associated pencil is $(\mu+\lambda\alpha)R+(\mu+\lambda\alpha)S=(\mu+\lambda\alpha)(R+S)=\hat{\lambda}\hat{S}$.  (ii)  By definition, the zero index number is the number of linearly independent constant vectors $\ket{v_i}$ such that $R\ket{v_i}=S\ket{v_i}=0$.  In this case, we must also have $\bar{R}\ket{\bar{v}_i}=\bar{S}\ket{\bar{v}_i} =0$.  It follows from \eqref{eq:reduced states} that $\rho_C\ket{\bar{v}_i}=0$ if and only if $\bar{R}\ket{\bar{v}_i}=\bar{S}\ket{\bar{v}_i} =0$ and since complex conjugation does not affect linear dependence, we have $rank(\rho_C)=n-g$.  An analogous argument shows that $rank(\rho_B)=m-h$.
\end{proof}

We now want to observe the effect of local invertible operators implemented by Alice, Bob, and Charlie; i.e. an SLOCC transformation.  Any such operation can be decomposed as $(A\otimes {\idmap}_{BC})({\idmap}_A\otimes B\otimes C)$ where Bob and Charlie first act, and then Alice follows alone.  When Bob and Charlie perform the invertible operator $B\otimes C$, it is easy to check that the transformation $\ket{R}_{BC}\to B\otimes C\ket{R}_{BC}$ corresponds to $R\to BRC^T$ and likewise for $S$.  Thus, the action of Bob and Charlie initiates the matrix pencil transformation $\mu R+\lambda S\to B(\mu R +\lambda S)C^T$.  In other words, \textit{local invertible operators of Bob and Charlie map matrix pencils to strictly equivalent ones}.

Any invertible operation by Alice can be represented by a matrix $\bigl(\begin{smallmatrix} a&b\\ c&d \end{smallmatrix} \bigr)$
 with $ad-bc\not=0$.  Then the most general action by Alice will transform the state $\ket{\Psi}$ as:
\begin{equation}
\ket{0}_A\ket{R}_{BC}+\ket{1}_A\ket{S}_{BC}\to\ket{0}_A(a\ket{R}_{BC}+c\ket{S}_{BC})+\ket{1}_A(b\ket{R}_{BC}+d\ket{S}_{BC}).
\end{equation}
Hence, the corresponding pencil transformation is $\mu R+\lambda S\to(\mu a+\lambda b)R+(\mu c+\lambda d)S=\hat{\mu}R+\hat{\lambda}S$ where $\hat{\mu}=\mu a +\lambda b$ and $\hat{\lambda}=\mu c+\lambda d$.  

What concerns us is how the transformation $(\mu,\lambda)\to(\hat{\mu},\hat{\lambda})$ affects the elementary divisors and minimal indices of a given pencil.  For the latter, care must be taken since minimal indices are defined by the degree of polynomials in variables $\mu$ and $\lambda$.  Nevertheless, the following lemma shows minimal indices to be an SLOCC invariant in $2\otimes m\otimes n$ systems. 

\begin{lemma}
The minimal indices of a given pencil remain invariant under the action of Alice.
\end{lemma}
\begin{proof}[\bf Proof]
\label{lm:AMinIndice}
Under an invertible transformation $(\mu,\lambda)\to (\hat{\mu},\hat{\lambda})=(a\mu+ b\lambda,c\mu+d\lambda)$, a polynomial $r$-component vector $p(\mu,\lambda)=\sum_{i=0}^{m-1}\sum_{j=0}^{n-1}x_{ij}\mu^i\lambda^j$ is identically zero iff $p(\hat{\mu},\hat{\lambda})\equiv 0$.  To see this, we can introduce the standard basis $\{\mathbf{e}_k\}_{k=1...r}$ and consider $p(\mu,\lambda)$ as an $rmn$-component vector in the space spanned by basis $\mu^i\lambda^j\mathbf{e}_k$.  Then the transformation $(\mu,\lambda)\to (\hat{\mu},\hat{\lambda})$ induces a homomorphism on this space which thus cannot map any nonzero zero vector to zero.  Consequently, for any set of polynomial vectors $\{\mathbf{x}_i(\mu,\lambda)\}_{i=1...n}$ (a) $(\mu R+\lambda S)\mathbf{x}_i(\mu,\lambda)\equiv 0$ iff $(\hat{\mu}R+\hat{\lambda}S)\mathbf{x}_i(\hat{\mu},\hat{\lambda})\equiv 0$, and (b) $\{\mathbf{x}_i(\mu,\lambda)\}_{i=1...n}$ is linearly independent iff $\{\mathbf{x}_i(\hat{\mu},\hat{\lambda})\}_{i=1...n}$ is linearly independent, where linear independence means that for polynomials $\{p_i(\mu,\lambda)\}_{i=1...n}$, $\sum_{i=1}^np_i(\mu,\lambda)\mathbf{x}_i(\mu,\lambda)\equiv 0\Rightarrow p_i(\mu,\lambda)\equiv 0$ for all $i$.  Next, we claim that (c) for any set of linearly independent scalar vectors $\{x_{ij}\}_{j=0...\epsilon}$ with $x_{i\epsilon}\not=0$, the highest degree of $\lambda$ having a nonzero vector coefficient in $\mathbf{x}_i(\hat{\mu},\hat{\lambda})=\sum_{j=0}^\epsilon x_{ij}\hat{\mu}^{\epsilon-j}\hat{\lambda}^j$ is the same as that in $\mathbf{x}_i(\mu,\lambda)=\sum_{j=0}^\epsilon x_{ij}\mu^{\epsilon-j}\lambda^j$.  This follows because the coefficient of $\lambda^\epsilon$ in $\sum_{j=0}^\epsilon x_{ij}\hat{\mu}^{\epsilon-j}\hat{\lambda}^j=\sum_{j=0}^\epsilon x_{ij}(a\mu+b\lambda)^{\epsilon-j}(c\mu+d\lambda)^j$ is $\sum_{j=0}^\epsilon x_{ij}b^{\epsilon-j}d^j$ which is non-vanishing due to the linear independence of $\{x_{ij}\}_{j=0...\epsilon}$.  

From (a), (c) and the linear independence of $\{x_{ij}\}_{j=0...\epsilon_i}$ noted in the introductory discussion for any fundamental set of vectors, $\mathbf{x}_1(\mu,\lambda)$ is a minimum degree polynomial in the null space of $\mu R+\lambda S$ iff $\mathbf{x}_1(\hat{\mu},\hat{\lambda})$ is a minimum degree polynomial in the null space of $\hat{\mu} R+\hat{\lambda} S$.  Now suppose that $\{\mathbf{x}_i(\mu,\lambda)\}_{i=1...n}$ are the first $n$ vectors in a fundamental set for $\mu R+\lambda S$ iff $\{\mathbf{x}_i(\hat{\mu},\hat{\lambda})\}_{i=1...n}$ are the first $n$ vectors in a fundamental set for $\hat{\mu} R+\hat{\lambda} S$.  Then by (c), $\mu R+\lambda S$ and $\hat{\mu} R+\hat{\lambda} S$ will have the same first $n$ minimal indices.  From (a), (b) and (c) again, $\mathbf{x}_{n+1}(\hat{\mu},\hat{\lambda})$ will be the next vector in the same fundamental set for $\mu R +\lambda S$ iff $\mathbf{x}_{n+1}(\hat\mu,\hat\lambda)$ is likewise for $\hat{\mu} R+\hat{\lambda} S$.  Hence by induction and by running the exact same argument on $(\mu R+\lambda S)^T$, the lemma is proven. 

\end{proof}

As for the elementary divisors, the situation is more complex since Alice's transformation can induce a mixing between infinite and finite divisors. By direct substitution, it follows immediately that after normalization, the divisors transform as
\begin{equation}
\mu^{e_i'}\to\begin{cases}(\mu\tfrac{a}{b}+\lambda)^{e_i'}&\text{if}\;\;b\not=0\\ \mu^{e_i'}&\text{if}\;\;b=0,\end{cases}\hspace{.5cm}\text{and}\hspace{.5cm}(\mu x_i+\lambda)^{e_{ij}}\to\begin{cases}(\mu\tfrac{ax_i+c}{bx_i+d} +\lambda)^{e_{ij}}&\text{if}\;\; bx_i+d\not=0\\ \mu^{e_{ij}}&\text{if}\;\;bx_i+d=0.\end{cases}
\end{equation} 
We see that depending on the choice of $A$, infinite divisors can become finite and finite can become infinite.  More importantly, given any general state having finite elementary divisors $\{(\mu x_i+\lambda)^{e_{ij}}\}$, it is always possible for Alice to perform an invertible operation such that $a\not=0$ and $\{bx_i+d\not=0\}$ for all $i$.  As a result, we see that
\begin{equation}
\label{eq:infdivsimp}
\textit{Any matrix pencil is SLOCC equivalent to one having no infinite divisors}.
\end{equation}
This observation simplifies the following analysis considerably since the general problem of determining general SLOCC equivalence is reduced to the problem of equivalence among states having only finite elementary divisors.  Combining the previous observations with Lemma \ref{kron}, we arrive at the following theorem and a main result of this article.

\begin{theorem}
\label{thm:main}
Two $2\otimes m\otimes n$ states $\ket{\psi}$ and $\ket{\phi}$ having only finite elementary divisors $\{(\mu x_i+\lambda)^{e_{ij}}\}$ and $\{(y_i+\lambda)^{f_{ij}}\}$ respectively are SLOCC equivalent if and only if their corresponding pencils are of the same rank, have the same minimal indices, $e_{ij}=f_{ij}$ for all $i,j$, and there exists a linear fractional transformation (LFT) relating the $x_i$ and $y_i$; i.e. for all $i$
\begin{equation}
\frac{ax_i+c}{bx_i+d}=y_i\;\;\;(ad-bc\not=0).
\end{equation}
\end{theorem}

A nice property of LFTs is that given any two trios $\{x_1,x_2,x_3\}$ and $\{y_1,y_2,y_3\}$ each with distinct values, there always exists a \textit{unique} LFT relating the sets \cite{Brown-2004a}.  The form of the transformation is given by the determinants
\begin{equation}
\label{eq:LFT}
a=\begin{vmatrix}x_1y_1&y_1&1\\x_2y_2&y_2&1\\x_3y_3&y_3&1\end{vmatrix},\;\;\;b=\begin{vmatrix}x_1y_1&x_1&y_1\\x_2y_2&x_2&y_2\\x_3y_3&x_3&y_3\end{vmatrix},\;\;\;c=\begin{vmatrix}x_1&y_1&1\\x_2&y_2&1\\x_3&y_3&1\end{vmatrix},\;\;\text{and}\;\;d=\begin{vmatrix}x_1y_1&x_1&1\\x_2y_2&x_2&1\\x_3y_3&x_3&1\end{vmatrix}.
\end{equation}

We are now able to present an algorithm for determining whether two general $2\otimes m\otimes n$ pure states $\ket{\psi}$ and $\ket{\phi}$ are SLOCC equivalent.
\begin{itemize}
\item[(I)] Input pencils $\mathcal{P}_{\psi}$ and $\mathcal{P}_{\phi}$ and determine their rank, minimal indices and elementary divisors.  As noted above, this step can be achieved via polynomial-time algorithms.  If the rank or minimal indices are not the same, $\ket{\psi}$ and $\ket{\phi}$ are inequivalent.  Otherwise, perform an arbitrary LFT on them so the pencils only have finite elementary divisors $\{(\mu x_i+\lambda)^{e_{ij}}\}$ and $\{\mu y_i+\lambda\}^{f_{ij}}$ respectively.  By observation \eqref{eq:infdivsimp} this can always be done.
\item[(II)]  Fix any three distinct $x_i$ corresponding to divisors of powers $e_{ij}$.  Choose any sequence of three distinct $y_i$ whose corresponding powers satisfy $f_{ij}=e_{ij}$ and determine the LFT relating $(x_1,y_1)$, $(x_2,y_2)$, and $(x_3,y_3)$ according to \eqref{eq:LFT}.  Choose a new $x_i$ and determine if the LFT relates it to any remaining $y_i$ belonging to an elementary divisor of the same power.  By uniqueness of the LFT, if there is no such $y_i$, the states are not equivalent.  If there is, choose another $x_i$ and repeat the search on the remaining $y_i$.
\item[(III)]  If a perfect matching exists for all $x_i$ and $y_i$, then the states are equivalent.  If not, repeat step (II) by choosing another ordered trio of the $y_i$.  If no LFT exists for all possible trios, the states are not equivalent.
\end{itemize}

The Kronecker canonical form of an $m\times n$ pencil can be computed
in time $O(m^2n)$ (see the algorithm by Beelen and Van Dooren~\cite{Beelen}).
For sets of $t$ elementary divisors, Step (II) this algorithm will require at most $O(t^3)$ steps.  Thus
the total running time is $O(m^2n + \min\{m, n\}^3)$.
Furthermore, the algorithm is constructive in nature because if two states are SLOCC equivalent, we determine the the specific $a,b,c,d$ constituting Alice's operator in the transformation $\ket{\phi}\SLOCCprec\ket{\psi}$.  The operators Bob and Charlie are to perform can be determined from the invertible matrices that bring pencils $\mathcal{P}_\psi$ and $\mathcal{P}_\phi$ to their canonical forms of \eqref{eq:kron} and are so-obtained by a Gaussian elimination procedure \cite{Gantmacher-1959a}.  Hence, not only does our algorithm determine whether two states are equivalent, but it provides the necessary operators achieving the transformations.   

%\bigskip\noindent
\section{All Tripartite Systems with a Finite SLOCC Equivalence Partitioning}
%\bigskip

To count and characterize all the orbits, we will essentially find what combination of minimal indices and elementary divisors fit in an $m\times n$ matrix of form \eqref{eq:kron}.  A few simplifications will assist in this process.  First, since any $m\times n$ pencil is simply the matrix transpose of an $n\times m$ one, it is enough to just consider $m\leq n$.  Next, for a given dimension, we must only study the equivalence classes with Bob and Charlie having maximal local ranks since any rank deficient case will correspond to a class of maximum local ranks in a smaller dimension.  To this end, Theorem \ref{thm:localranks} allows us to immediately determine the local ranks associated with each equivalence class.  Furthermore, as evident from the Schmidt decomposition of any state with respect to bipartition AB:C, Charlie's local rank cannot exceed the product of Alice and Bob's.  Consequently, if $n\geq 2m$, any state of a $2\otimes m\otimes n$ system is the same as one in a $2\otimes m\otimes 2m$ system up to a local change of basis on Charlie's part.  This means that for the task of finite enumeration, we only need to consider systems up to dimensions $2\otimes 2\otimes 4$ and $2\otimes 3\otimes 6$.     

One further property of each equivalence class that we are able to study is the tensor rank.  The tensor rank of a state is the minimum number of product states whose linear span contains the state, and this quantity turns out to be invariant under invertible SLOCC transformations \cite{Dur-2000a}.  For bipartite systems, the tensor rank is equivalent to the Schmidt rank, and a non-increase in Schmidt rank is also a sufficient condition for SLOCC convertibility between two such states; SLOCC equivalence classes are characterized completely by the Schmidt rank.  Interestingly, in three qubit systems, tensor rank is also sufficient to distinguish between the various equivalence classes.  However, we find that even for systems having a finite partitioning, the tensor rank is an insufficient measure for determining SLOCC equivalence.  Our results follow from previous research on the tensor rank of matrix pencils done by Ja' Ja' \cite{Ja'Ja'-1978a} and rederived in Ref. \cite{Burgisser-1997a}.
\begin{lemma}{\upshape \cite{Ja'Ja'-1978a},\cite{Burgisser-1997a}}
Let $\mathcal{P}_{(R,S)}$ be a pencil with no infinite divisors in canonical form \eqref{kron} with minimal indices $\epsilon_1,...,\epsilon_p$ and $\nu_1,...,\nu_q$ and $J$ an $l\times l$- sized pencil.  Furthermore, let $\delta(J)$ denote the number of invariant polynomials containing at least one nonlinear elementary divisor.  Then the tensor rank of $\mathcal{P}_{(R,S)}$ is given by
\begin{equation}
\label{eq:tensorrank}
\sum_{i=1}^p(\epsilon_i+1)+\sum_{j=1}^q(\nu_j+1)+l+\delta(J).
\end{equation}
\end{lemma}

A summary of all the equivalence classes is provided in Table \ref{tbl:Equivclass} in Appendix.  We see that there are 26 distinct SLOCC classes for $2\otimes 3\otimes n$ $(n\geq 6)$ systems.  This reproduces the findings of Chen \textit{et al.} \cite{Chen-2006a} here obtained in an entirely different way by using matrix pencil analysis.
%\bigskip\noindent
\section{Non-Invertible Transformations}
%\bigskip

A natural question is whether it is possible to transform from one class to another via non-invertible transformations.  One obvious constraint is that states with full local ranks cannot preserve their ranks under a non-invertible transformation.  Consequently, we cannot convert among the states belonging to the same dimensional grouping above.  A possible conjecture might be that unidirectional convertibility is achievable if none of the local ranks increase and at least one decreases; certainly three qubit systems satisfy this hypothesis.  This, however, is false in general as we will now observe.

Let $\ket{\psi}$ be some state having maximal local ranks of $(2,m,n)$ and suppose $\ket{\phi}$ is a state with ranks $(2,m,n-1)$.  If $\ket{\phi}\SLOCCprec\ket{\psi}$, Alice and Bob's matrices inducing the transformation will be full rank while Charlie's will have rank $n-1$.  As for the latter, any such operator can be decomposed into a series of elementary column operations on $\mathcal{P}_\psi$ followed by a mapping of the $n^{th}$ column to a linear combination of the first $n-1$; this $n-1$-columned subpencil corresponds to the target state $\ket{\phi}$.  >From the commutation relations of elementary operations, if we neglect permutations, $\ket{\phi}$ will be some $n-1$-column subset of the original pencil following the application of just column-multiplying and column-addition transformations.  Moreover, if column $i$ is the linearly dependent column, then immediately after all column-additions of the $i^{th}$ column are performed, the remaining $n-1$ columns must be equivalent to $\ket{\phi}$.  As a result, we obtain the following criterion.  

\begin{theorem}
\label{thm:noninvert}
Let $\ket{\psi}$ and $\ket{\phi}$ be states with local ranks $(2,m,n)$ and $(2,m,n-1)$, and let $c_1,...,c_n$ denote the columns of $\mathcal{P}_\psi(\mu,\lambda)$.  Then $\ket{\phi}\SLOCCprec\ket{\psi}$ iff for some $1\leq i\leq n$, there exists constants $a_1,..,a_{i-1},a_{i+1},...,a_n$ and some invertible linear transformation $(\mu,\lambda)\to(\hat\mu,\hat\lambda)$ such that the pencil $\mathcal{P}_{\psi_i}(\hat\mu,\hat\lambda)=[c_1+a_1c_i,...,c_n+a_nc_i]$ is equivalent to $\mathcal{P}_\phi(\mu,\lambda)$.
\end{theorem} 

In general, for transformations in which Charlie's rank decreases to $n-k$, one need only modify this theorem by considering subpencils of $\mathcal{P}_\psi$ having $n-k$ columns where to each of the columns is added a linear combination of the $k$ non-included columns.  Likewise, to account for transformations when Bob's local rank decreases, the above criterion can be applied with the analysis conducted on the rows of $\mathcal{P}_\psi(\mu,\lambda)$ instead of its columns.  

On the surface, Thm. \ref{thm:noninvert} has limited value since it involves a search for values $a_1$,..,$a_{i-1}$, $a_{i+1}$,...,$a_n$ over the complex numbers.  However, in many cases, it is easy to see whether or not such a collection of numbers can be found.  For example, for $1\leq i\leq 4$ in (ABC-19), upon any choice of the $a_j$ and any transformation $\hat\mu,\hat\lambda$, the resultant pencil $\mathcal{P}_{\psi_i}(\hat\mu,\hat\lambda)$ will either be rank two or it will have an elementary divisor of degree at least one.  However, the state (ABC-18) is rank three with no non-trivial elementary divisors.  Thus, the transformation (ABC-19)$\to$(ABC-18) is impossible.  On the other hand, for the state (ABC-17), when $i=1$, we have $\det\mathcal{P}_{\psi_1}(\mu,\lambda)=\lambda[\lambda^2-\mu(\tfrac{a_2}{a_3}\lambda+\tfrac{1}{a_3}\mu)]$ for $a_3\not=0$.  The state (ABC-8) has $\det\mathcal{P}_\phi(\mu,\lambda)=\lambda(\mu+\lambda)(2\mu+\lambda)$.  By choosing $c_2=\tfrac{3}{2}$ and  $c_3=-\tfrac{1}{2}$, these polynomials become equal as well as the elementary divisors, the ranks, and the minimal indices of the pencils.  Thus (ABC-17)$\to$(ABC-8) is achievable by SLOCC.

In a manner similar to that just described, we have used Thm. \ref{thm:noninvert} to analyze all possible transformations among the $2\otimes 3\otimes n$ equivalence classes.  Figure~(\ref{fig:chart}) in Appendix depicts the SLOCC hierarchy among the classes.

%\bigskip\noindent
\section{Conclusions and Future Research}
%\bigskip

In this article, we have used the theory of matrix pencils to study $2\otimes m\otimes n$ pure quantum states.  In doing so, we were able to derive a polynomial time algorithm for deciding SLOCC equivalence of such states.  For all tripartite systems having a finite number of equivalence classes, we have obtained state representatives and determined the partial ordering among these classes based on a criterion for general SLOCC convertibility in $2\otimes m\otimes n$ systems.   It is interesting to note that in the hierarchy chart of Fig. 2, there exists certain transformations that are impossible even though the local rank of Charlie decreases by two.  The transformation (ABC-14) to (ABC-7) is such example.

A natural extension of this work is to find efficient algorithms for deciding LOCC equivalence, LOCC convertibility,
and SLOCC convertibility in $2\otimes m\otimes n$ systems. We have made progress on those questions.
Another natural next line of inquiry might to consider $p\otimes m\otimes n$ systems and their corresponding degree $p$ matrix polynomials.  Indeed, much research has been conducted on higher degree elements, especially those having special properties such as being symmetric \cite{Gohberg-2009a}.  Unfortunately, there exists no corresponding characterization like Kronecker's for strict equivalence of matrix pencils of degree greater than two.  Making the project of generalizing to higher degrees more dubious is the fact that determining SLOCC equivalence for $p\otimes m\otimes n$ can be reduced from a tensor rank calculation on a set of $p$ bilinear forms \cite{Chitambar-2008a}, and this problem has no known solution for $p>2$ \cite{Burgisser-1997a} (the general problem is, in fact, NP-Hard \cite{Haastad-1990a}). 

As noted in the introduction, we are not the first to study SLOCC convertibility in multipartite systems, and it would be interesting to try and develop the relationship between our results and the work of others.  For example, Miyake's results involve ``hyperdeterminants'' and their singularities \cite{Miyake-2003a}.  It would be valuable to investigate the correspondence between matrix pencils and hyperdeterminants or to introduce the connection to the quantum information community if such a correspondence has already been obtained.  In another work, Liang \textit{et al.} have recently proven a set of conditions both necessary and sufficient for the convertibility of two qubit mixed bell-diagonal states \cite{Liang-2008a}.  As these mixed states can be considered pure with respect to a $2\otimes 2\otimes 4$ system, it would be fruitful to study transformations between tripartite ``purified'' bell diagonal states via our matrix pencil construction and compare it to the convertibility conditions in Ref. \cite{Liang-2008a}.  Doing so might suggest ways in which purified tripartite pencils can assist in deciding equivalence between general $2\otimes n$ mixed states.

\bibliographystyle{abbrv}

\bibliography{QuantumBib} 

\newpage
\appendix
\section*{Appendix A}
\begin{longtable}[ht]{|c|c|c||c|c|c|}\hline
\label{tbl:Equivclass}
 Representative & Local Ranks & Tensor Rank &Representative & Local Ranks & Tensor Rank\\\hline
 (A:B:C) & (1,1,1) & 1 &
  (AB:C) & (2,2,1) & 2 \\\hline
  (AC:B) & (2,1,2) & 2 &
  (A:BC-1) & (1,2,2) & 2 \\\hline
  (ABC-1) & (2,2,2) & 2&
  (ABC-2) & (2,2,2) & 3\\\hline
  (ABC-3) & (2,2,3) & 3&
  (ABC-4) & (2,2,3) & 3\\\hline
  (ABC-5) & (2,2,4) & 4&
  (ABC-6) & (2,3,2) & 3\\\hline
  (ABC-7) & (2,3,2) & 3&
  (ABC-8) & (2,3,3) & 3\\\hline
  (ABC-9) & (2,3,3) & 3&
  (A:BC-2) & (1,3,3) & 2\\\hline
  (ABC-10) & (2,3,3) & 4&
  (ABC-11) & (2,3,3) & 4\\\hline
  (ABC-12) & (2,3,3) & 4&
  (ABC-13) & (2,3,3) & 4\\\hline
  (ABC-14) & (2,3,4) & 4&
  (ABC-15) & (2,3,4) & 4\\\hline
  (ABC-16) & (2,3,4) & 5&
  (ABC-17) & (2,3,4) & 4\\\hline
  (ABC-18) & (2,3,4) & 4&
  (ABC-19) & (2,3,5) & 5\\\hline
  (ABC-20) & (2,3,5) & 5&
  (ABC-21) & (2,3,5) & 6\\\hline
  \caption{Equivalence Classes in $2\otimes 3\otimes 6$ Systems}
\end{longtable}

\begin{figure}[H]
\centering
\includegraphics[width=6in]{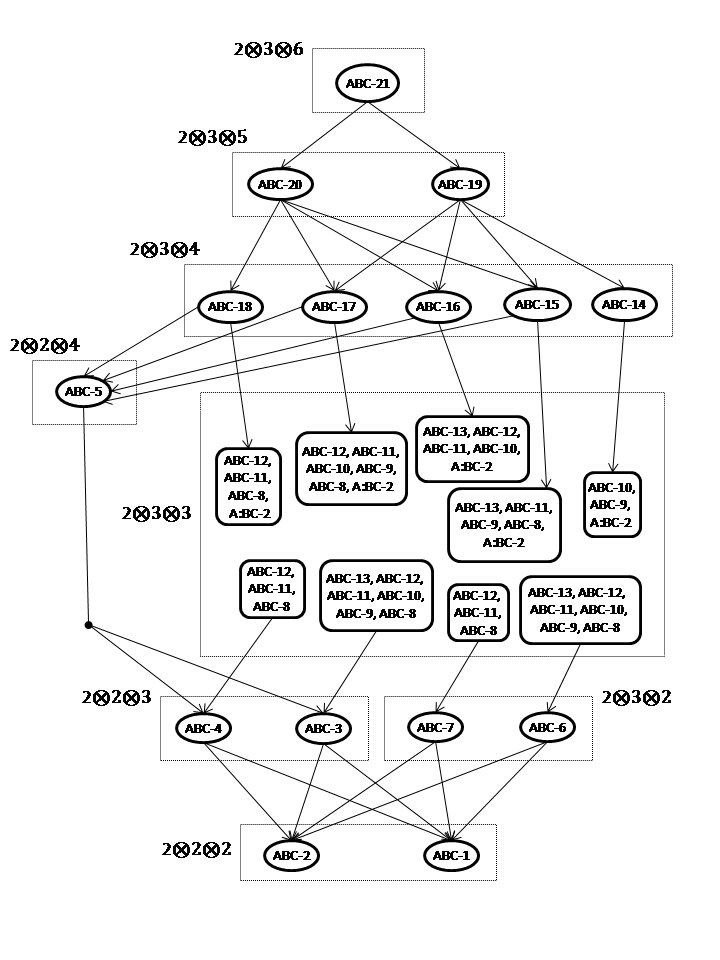}
\caption{Complete Hierarchy of SLOCC Equivalence Classes; Arrows Indicate a Non-invertible Transformation.}
\label{fig:chart}
\end{figure}

\end{document}